\documentclass[aps,prl,superscriptaddress,nofootinbib,twocolumn]{revtex4-2} 
\usepackage{amsfonts,amsmath,amssymb}
\usepackage{physics}
\usepackage[utf8]{inputenc}
\usepackage{caption}
\usepackage{subcaption}
\usepackage{graphicx}
\usepackage{hyperref}
\usepackage{xcolor}
\usepackage{siunitx}
\usepackage{tabularx}
\include{macros}
\begin{document}


\title{Zero- and finite-temperature electromagnetic strength distributions in closed- and open-shell nuclei from first principles}

\author{Y. Beaujeault-Taudi\`ere}
\affiliation{CEA, DAM, DIF, 91297 Arpajon, France}
\affiliation{Universit\'e Paris-Saclay, CEA, Laboratoire Mati\`ere en Conditions Extr\^emes, 91680 Bruy\`eres-le-Ch\^atel, France}

\author{M. Frosini}
\affiliation{CEA, DEN, IRESNE, DER, SPRC, 13108  Saint-Paul-l\`es-Durance, France}
\affiliation{IRFU, CEA, Universit\'e Paris-Saclay, 91191 Gif-sur-Yvette, France}

\author{J.-P. Ebran}
\affiliation{CEA, DAM, DIF, 91297 Arpajon, France}
\affiliation{Universit\'e Paris-Saclay, CEA, Laboratoire Mati\`ere en Conditions Extr\^emes, 91680 Bruy\`eres-le-Ch\^atel, France}

\author{T. Duguet}
\affiliation{IRFU, CEA, Universit\'e Paris-Saclay, 91191 Gif-sur-Yvette, France}
\affiliation{KU Leuven, Department of Physics and Astronomy, Instituut voor Kern- en Stralingsfysica, 3001 Leuven, Belgium}

\author{R. Roth}
\affiliation{Institut f\"ur Kernphysik, Technische Universit\"at Darmstadt, 64289 Darmstadt, Germany}
\affiliation{Helmholtz Forschungsakademie Hessen f\"ur FAIR, GSI Helmholtzzentrum, 64289 Darmstadt, Germany}

\author{V. Som\`a}
\affiliation{IRFU, CEA, Universit\'e Paris-Saclay, 91191 Gif-sur-Yvette, France}

\date{\today}

\begin{abstract}
    Ab initio approaches to the nuclear many-body problem have seen their reach considerably extended over the past decade. However, collective excitations have been scarcely addressed so far due to the prohibitive cost of solving the corresponding equations of motion. Here, a numerically efficient method to compute electromagnetic response functions at zero- and finite-temperature in superfluid and deformed nuclei from an ab initio standpoint is presented and applied to $^{16}$O, $^{28}$Si, $^{46}$Ti and $^{56}$Fe. This work opens the path to systematic ab initio calculations of nuclear responses to electroweak probes across a significant portion of the nuclear chart. 
\end{abstract}

\maketitle


\paragraph{Introduction. --} As strongly correlated many-body systems, atomic nuclei display complex behaviours, from deformation~\cite{ShapeQPT,Shapecoex} to superfluid and molecular instabilities~\cite{Ebran2012,EbranQPT,Clusterreview}. The interference of these phenomena gives birth to a vast diversity of possible arrangements of the nucleons inside nuclei, which eventually imprint specific signatures in ground- and excited states properties. In this context, the nuclear response to electroweak probes, in addition to providing key inputs to various applications, e.g. to reaction mechanisms involved in nucleosynthesis processes, such as p- and r-processes~\cite{ARNOULD200797}, is a valuable tool to scrutinize nucleonic correlations.

Against this background, the theoretical description of nuclear systems based
on a web of interlocking effective field theories (EFTs), among which ab initio approaches occupy a privileged position, has become a dynamic and productive area of research in recent years~\cite{Bontems2021}. Significant progress in the ab initio treatment of nuclear systems has been driven by (i) the construction of consistent and systematically improvable nuclear interactions within the frame of chiral EFT ($\chi$EFT)~\cite{wei79,ent17,chipot}, and (ii) the formulation of new and refined  many-body schemes with controlled uncertainties~\cite{heikoguidedtour,frosiniI,frosiniII,frosiniIII}. However, in spite of the ever-increasing applicability of such methods across the table of nuclides~\cite{Coraggio21}, the ab initio study of strength distributions for electromagnetic transition
operators largely remains uncharted territory.

Early attempts to compute electromagnetic response functions from an ab initio standpoint made use of the random phase approximation (RPA)~\cite{paa06} and its quasiparticle extension (QRPA)~\cite{her11}.
More recent calculations have been performed either via coupled cluster (CC)~\cite{Bacca:2014rta,Miorelli:2016qbk,Miorelli:2018dcb,Bonaiti:2021kkp}, no-core shell model (NCSM)~\cite{Stumpf2017} or Dyson self-consistent Green's function (SCGF)~\cite{Raimondi:2019} approaches. However, calculations have been mostly limited to doubly closed-shell systems so far. The extension of such methods to singly and doubly open-shell nuclei is highly non-trivial because of the associated increase in numerical cost. As a result, a first-principles computation of the nuclear response to electroweak probes is currently missing for the large majority of nuclei, even in the medium-mass sector. 

Eventually, the description of collective excitations has been the hallmark of phenomenological models so far, especially that of energy density functionals (EDFs) based on the RPA and its extensions~\cite{EDFebook}. In this context, a novel scheme to solve (Q)RPA equations was proposed a few years ago in Refs.~\cite{nak07,avo11}. This approach, coined as the (quasiparticle) finite amplitude method ((Q)FAM), replaces the intensive calculation and diagonalization of the QRPA matrix by a set of non-linear equations of similar dimension to that of the static Hartree-Fock-Bogoliubov (HFB) mean-field approach it builds upon. The QFAM has proven to be a very efficient tool to obtain electric~\cite{ina09,kor11,kor15,ois16} and charge-exchange~\cite{mus14,eng20} strength functions, as well as to determine collective inertia \cite{hin15_coll_intertia,was21}, quasiparticle-vibration coupling \cite{lit21}, discrete eigenmodes~\cite{hin13} and sum rules~\cite{hin15_sum_rules}. 

On the ab initio side, some efforts have been recently dedicated to the calculation of ground and low-lying excited states in doubly open-shell, e.g. deformed and superfluid, nuclei~\cite{Novario:2020kuf,Yao20,frosiniI,frosiniII,frosiniIII,Hagen:2022tqp}. These developments are well suited to be complemented with QFAM-type algorithms to access higher-lying collective excitations. The present letter thus reports on the first zero-temperature (ZT) and finite-temperature (FT) electromagnetic strength distributions computed in doubly open-shell nuclei from first principles. Based on the QFAM, resulting HFB-based QRPA (HFB-QRPA) calculations (i) employ full two- and three-nucleon interactions rooted into quantum chromodynamics (QCD) via $\chi$EFT, (ii)  apply indistinctly to doubly closed-shell, singly open-shell and doubly open-shell nuclei, (iii) displays a favourable scaling with mass number.

Thereafter, the ab initio HFB-QRPA is first benchmarked in the doubly closed-shell nucleus $^{16}$O based on the ZT electric isovector dipole (IVD) $E1$ photoabsorption cross section. The same observable is confronted to experimental data in doubly open-shell $^{28}$Si and $^{46}$Ti nuclei before investigating the role of the temperature in $^{56}$Fe. Details about the FT HFB-QRPA formalism through the FT-QFAM, a numerical benchmark against results obtained via the diagonalization method, the employed $\chi$EFT Hamiltonians and the numerical settings are provided as Supplemental Materials.  

\paragraph{Benchmarks in $^{16}$O. --} Ab initio methods aim at accessing approximate  solutions of the many-body Schr\"odinger equation that are systematically improvable. Present ab initio calculations employ the QRPA based on an HFB reference state\footnote{The QRPA based on an HFB state reduces to the RPA based on a Hartree Fock (HF) reference state in closed-shell nuclei.}~\cite{rin80}, which a priori constitutes a rather severe approximation that is not guaranteed to deliver converged enough results. However, in the following, the fact that QRPA targets differential quantities, i.e. excitation energies and associated transition probabilities, is proven to largely benefit from the cancellation of so-called {\it dynamical} correlations consistently added to ground- and excited states\footnote{Such a quantitative cancellation was also recently demonstrated to occur for low-lying collective states computed via the projected generator coordinate method perturbation theory (PGCM-PT)~\cite{frosiniI,frosiniII,frosiniIII}. Specifically, spectroscopic properties delivered by the PGCM itself, a method intimately related to QRPA~\cite{JANCOVICI1964,BRINK1968}, were shown to be essentially identical to those obtained after the addition of dynamical correlations to both ground and low-lying excited states.} on top of HFB-QRPA. 

\begin{figure}[htb!]
    \centering
    \includegraphics[width=1.1\linewidth]{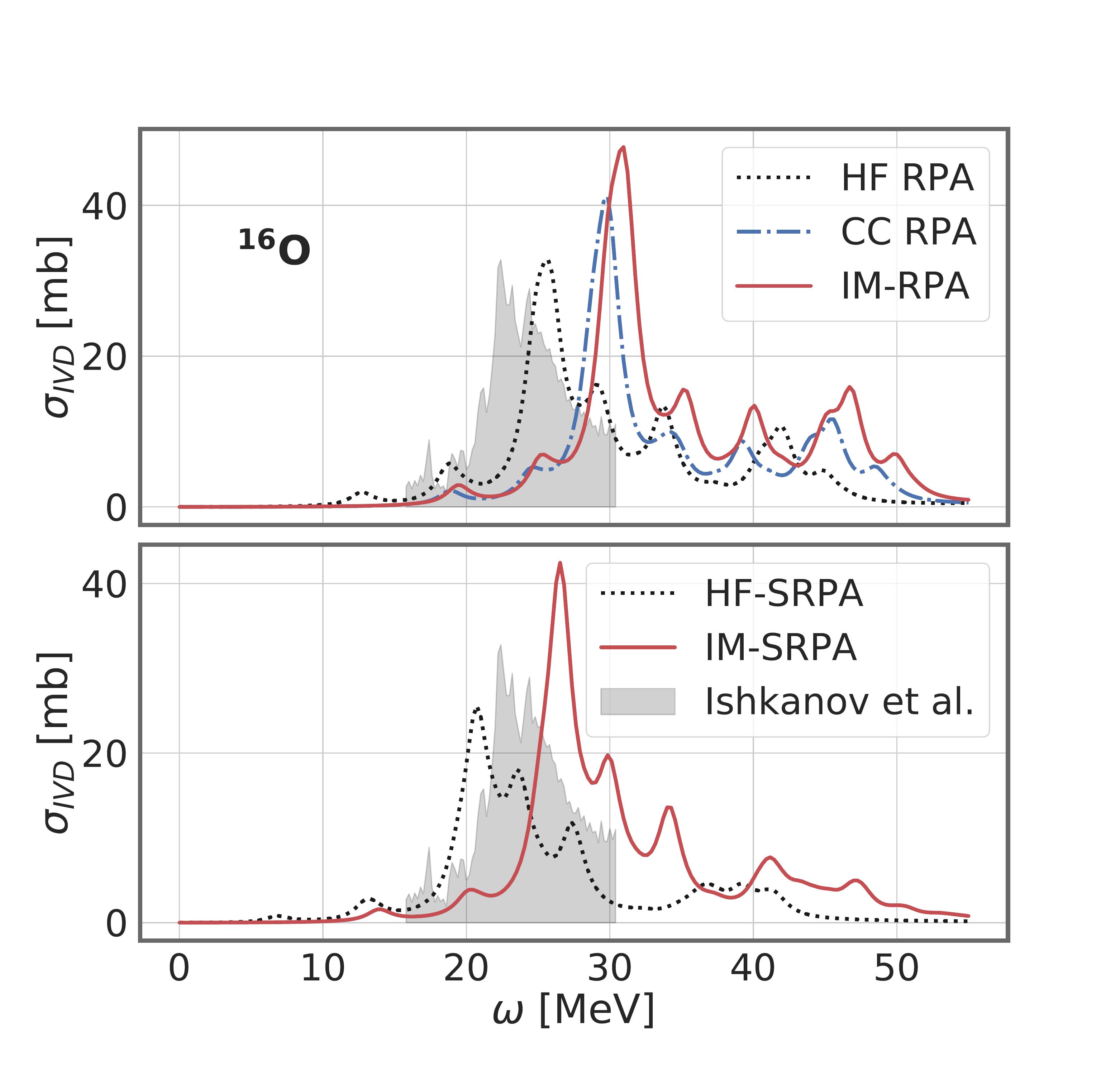}
    \caption{Integrated isovector E1 photoabsorption cross section of $^{16}$O computed as a function of the excitation energy $\omega$ through HF-RPA, CC-RPA and IM-RPA calculations (upper panel) as well as through HF-SRPA and IM-SRPA calculations (bottom panel). The NNLO$_\text{sat}$ chiral Hamiltonian~\cite{nnlosat}  softened through a SRG transformation ($\alpha = 0.08$\,fm$^4$) is employed whereas  experimental data are taken from Ref.~\cite{Ish}.}  
    \label{fig:O16a}
\end{figure}

To illustrate this key result, the ZT electric IVD photoabsorption cross section computed via HF-RPA in the doubly-closed shell $^{16}$O nucleus is compared in Fig.~\ref{fig:O16a} to the same quantity computed via coupled-cluster RPA (CC-RPA) and in-medium RPA (IM-RPA) in the upper panel as well as via HF second RPA (HF-SRPA) and IM-SRPA in the bottom panel. More details on these hybrid methods can be found in Refs.~\cite{trippel2016,roth2023}. All calculations are performed with the state-of-the-art next-to-next-to-leading-order (N$^2$LO) NNLO$_\text{sat}$ chiral Hamiltonian~\cite{nnlosat} softened through a similarity renormalization group (SRG) transformation characterized by a flow parameter $\alpha = 0.08$\,fm$^4$. While CC-RPA and IM-RPA constitute two alternative ways to add dynamical correlations to the ground state~\cite{trippel2016}, SRPA does the same for the excited states accessed via the E1 photoabsorption process. 

One first observes in the upper panel that explicitly correlating the ground-state mechanically lowers the ground-state energy of $^{16}$O by $5-6$\,MeV and thus consistently shifts the excitation strength upward by the same amount, without changing significantly the overall profile. While operating very differently, CC-RPA and IM-RPA deliver consistent results in this respect\footnote{A consistent effect is obtained via {\it dressed} RPA calculations based on self-consistent Green's function theory, where dressing one-nucleon propagators relates to using a correlated rather than a HF ground-state~\cite{Raimondi:2019}.}. Contrarily, correlating the excited states via  SRPA systematically pushes the strength downward by about 5\,MeV. When correlating both ground and excited-states at once via IM-SRPA, a remarkable compensation effect is observed. Consequently, the IM-SRPA IVD photoabsorption cross section is very similar to the initial HF-RPA one, even though not in perfect agreement for the energy position of the main peaks and for the intensity of the associated strength~\cite{carter22a}. As visible from Tab.~\ref{alphaD}, a sensitive quantity such as the dipole polarizability is impacted by dynamical correlations beyond HF-RPA~\cite{Miorelli:2018dcb}. In any case, the above analysis demonstrates that the basic ab initio HF-RPA quantitatively captures the main characteristics of the converged electromagnetic response in $^{16}$O.  

Because adding dynamical correlations to both ground- and excited states comes at a high computational cost, the validation of the HF(B)-(Q)RPA as a viable approach to collective excitations in nuclei constitutes a critical result. Indeed, it opens the possibility to perform systematic ab initio calculations of nuclear responses across a large portion of the nuclear chart. Based on the above analysis, all results presented below are obtained from ZT and FT HF(B)-(Q)RPA calculations.

\begin{figure}[htb!]
    \centering
    \includegraphics[width=\linewidth]{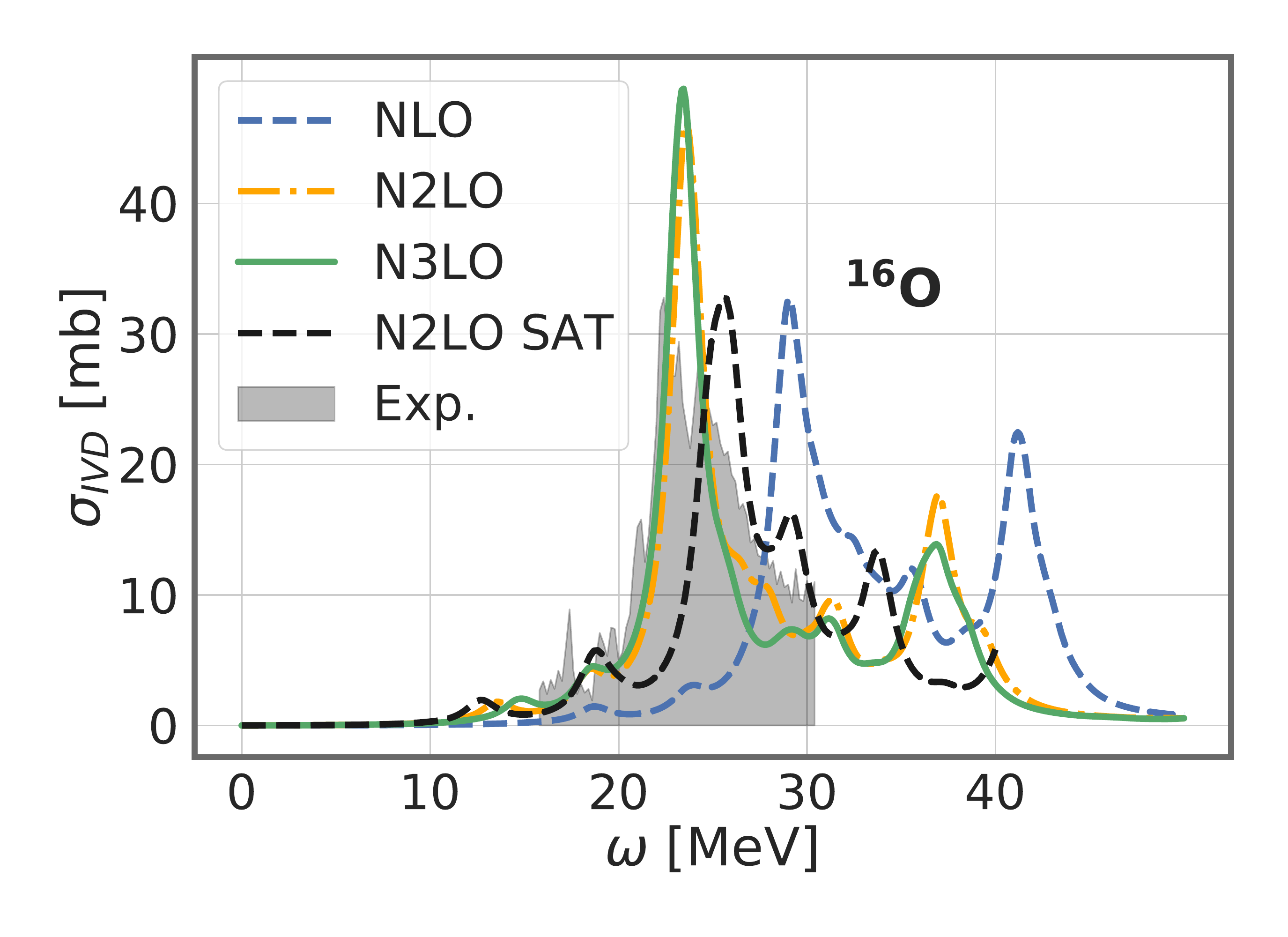}
    \caption{Integrated isovector E1 photoabsorption cross section of $^{16}$O calculated as a function of the excitation energy $\omega$ through HF-RPA. Results obtained with a consistent family of chiral Hamiltonians at NLO, N$^2$LO and N$^3$LO~\cite{HUTHER2020135651} are displayed in addition to results obtained from the NNLO$_\text{sat}$ Hamiltonian~\cite{nnlosat} already employed in Fig.~\ref{fig:O16a}. Experimental data are taken from Ref.~\cite{Ish}.}  
    \label{fig:O16b}
\end{figure}

A second important point relates to the uncertainty of nuclear responses associated with the employed chiral Hamiltonian. Fig.~\ref{fig:O16b} displays the $^{16}$O ZT isovector E1 photoabsorption cross section obtained from a consistent family of chiral Hamiltonians produced at NLO, N$^2$LO and N$^3$LO~\cite{HUTHER2020135651} and softened through a SRG transformation ($\alpha = 0.08$\,fm$^4$). Going from NLO to N$^2$LO, the strength is shifted down by about 6\,MeV whereas the height of the main peak is increased by nearly 50$\%$, thus bringing the calculation in agreement with experimental data~\footnote{It has been explicitly checked in $^{16}$O that the quality of the E1 IVD results is representative of the one obtained for other multipolarities of the transition operator.}. This is consistent with the systematic effect of three-nucleon interactions~\cite{trippel2016} that first contribute at N$^2$LO. Going from N$^2$LO to N$^3$LO, the cross section remains essentially unchanged, demonstrating an excellent convergence with respect to the Hamiltonian expansion. Fig.~\ref{fig:O16b} also shows the results displayed in Fig.~\ref{fig:O16a} for the NNLO$_\text{sat}$ Hamiltonian~\cite{nnlosat}. One observes a non-negligible shift with respect to the N$^2$LO results, the location of the main peak being shifted up by 1.5\,MeV and its strength being reduced by about 30$\%$. This highlights the remaining uncertainty associated with the way the low-energy constants entering the chiral Hamiltonian are adjusted on experimental data and with the way the Hamiltonian is regularized; see e.g. Ref.~\cite{LENPIC:2022cyu} for a recent account of these issues. Based on the above analysis, all results presented below are obtained with the N$^3$LO chiral Hamiltonian of Ref.~\cite{HUTHER2020135651}.

\begin{table}
\centering
\renewcommand{\arraystretch}{1.4}
\begin{tabular}{|l|c|c|c|c|}
\cline{1-5}
\multicolumn{1}{|c|}{$\alpha_{\text{D}}$} & \multicolumn{2}{|c|}{NNLO$_\text{sat}$} & \multicolumn{1}{|c|}{N$^3$LO} & \multicolumn{1}{|c|}{Exp}\\
\cline{2-4}
[fm$^3$] & HFB-QRPA & IM-SRPA &  HFB-QRPA  &  \\
\cline{2-4}
\noalign{\vskip\doublerulesep
         \vskip-\arrayrulewidth}
\hline
\multicolumn{1}{|l||}{$^{16}$O} & 0.63 & 0.53 & 0.61 & 0.58(1) \\
\hline
\multicolumn{1}{|l||}{$^{28}$Si} & - & - & 1.33  & - \\
\hline
\multicolumn{1}{|l||}{$^{46}$Ti} & - & - & 2.50 & - \\
\hline
\multicolumn{1}{|l||}{$^{56}$Fe} & - & - & 3.05 & - \\
\hline
\end{tabular}
\vspace{.35cm}
\caption{ZT dipole polarizability $\alpha_{\text{D}}$ in the nuclei of present interest. The experimental value in $^{16}$O is taken from Ref.~\cite{arhens75a}.}
\label{alphaD}
\end{table}

\paragraph{ZT IVD excitations in doubly open-shell nuclei. --} Dynamical correlations were shown above to be essentially identical in ground and excited states,  thus largely cancelling out in the $^{16}$O IVD photoabsorption cross section.  In addition, such correlations vary smoothly with mass number~\cite{trippel2016}. Contrarily, the magnitude of so-called {\it static} correlations, responsible for non-trivial collective behaviours such as superfluidity and deformation, evolves quickly with nucleon number, i.e. with the open- versus closed-shell character of the nucleus under consideration\footnote{A closed-shell nucleus such as $^{16}$O does not display significant static correlations.}. Furthermore, static correlations are known to {\it qualitatively} impact electromagnetic strength distributions such that their inclusion from the outset is mandatory. One efficient way to do so relies on the concept of spontaneous symmetry breaking~\cite{frosiniI,Hagen:2022tqp,EDFebook}, thus leading in the present context to the deformed HFB-QRPA extension of the spherical HF-RPA employed so far in ab initio calculations. While this extension is prohibitive for ab initio calculations when employing the traditional diagonalization method, the FAM language enables a  computationally affordable formulation of the deformed HFB-QRPA, tremendously enlarging the pool of accessible nuclei from few doubly-closed shell nuclei to thousands of singly and doubly open-shell ones.

\begin{figure}[htb!]
     \centering
        \includegraphics[width=\linewidth]{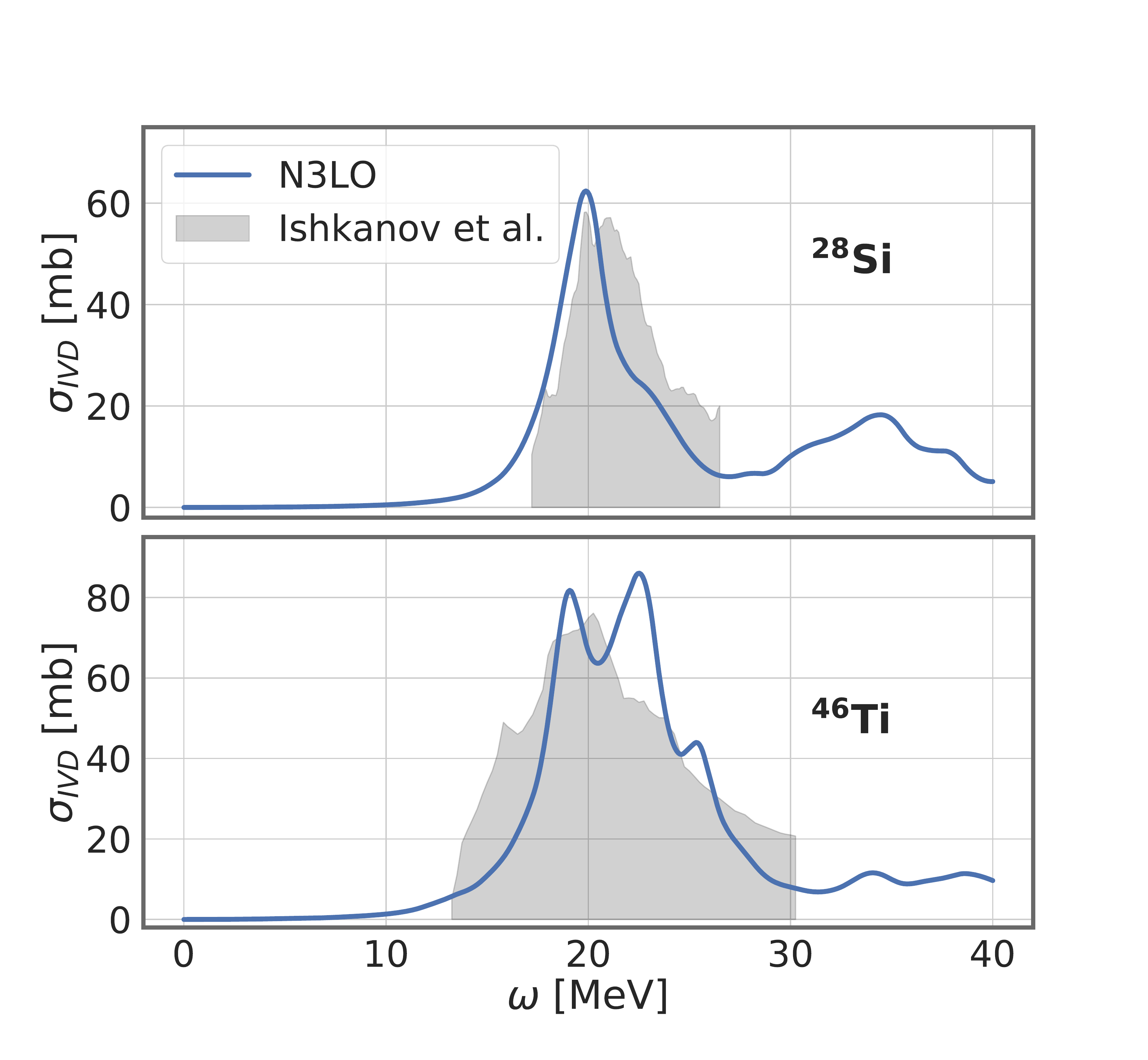}
        \caption{Integrated isovector E1 photoabsorption cross section of $^{28}$Si (upper panel) and $^{46}$Ti (lower panel) as a function of the excitation energy $\omega$. Experimental data are taken from Ref.~\cite{Ish}.}
        \label{fig:E1SiTi}
\end{figure}

Having validated the ZT HF-RPA in $^{16}$O, its deformed HFB-QRPA extension is now employed for doubly open-shell $^{28}$Si and $^{46}$Ti nuclei whose integrated isovector E1 photoabsorption cross sections are displayed in Fig.~\ref{fig:E1SiTi}. At the HFB level, the ground states of $^{28}$Si and $^{46}$Ti spontaneously break both U(1) and SO(3) symmetries, respectively associated with particle-number and angular-momentum conservation. In particular, the intrinsic deformation of $^{28}$Si ($^{46}$Ti) is found to be oblate (prolate).

In $^{28}$Si, the energy of the experimental isovector giant dipole resonance (GDR) at \SI{20}{\mega\electronvolt} and of the shoulder appearing around \SI{26}{\mega\electronvolt} are accurately reproduced. Similar conclusions can be drawn for $^{46}$Ti, even if the width of the resonance is slightly underestimated in spite of delivering the three-peak structure. A better detailed reproduction would benefit from an explicit inclusion of dynamical correlations beyond HFB-QRPA that would further fragment the strength. In any case, the present examples demonstrate the capacity of the ab initio HFB-QRPA to capture decisive static correlations in (doubly) open-shell nuclei and to be consistent with experimental data.

\begin{figure}[t!]
    \centering
    \includegraphics[width=1\linewidth]{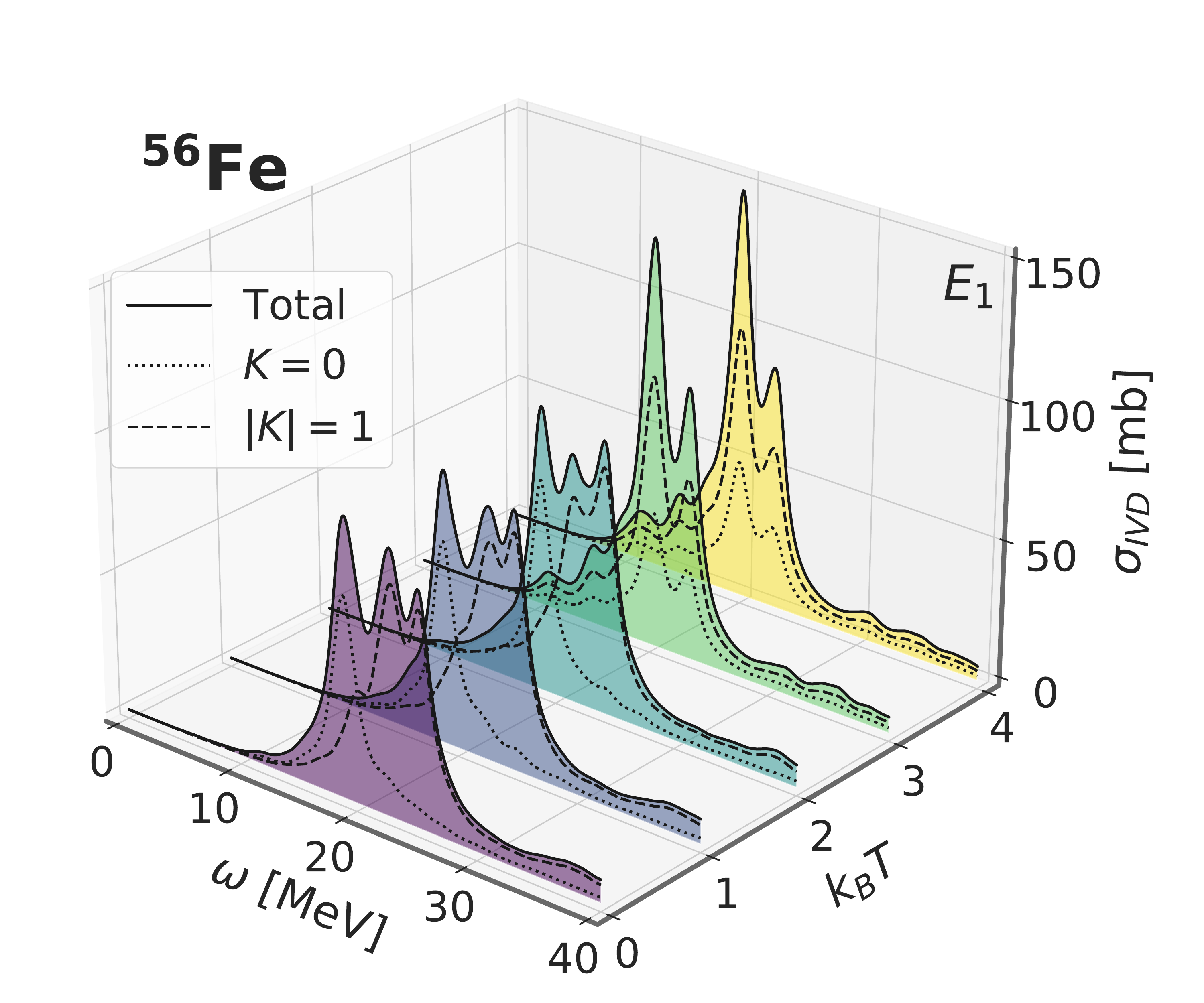}
    \includegraphics[width=1\linewidth]{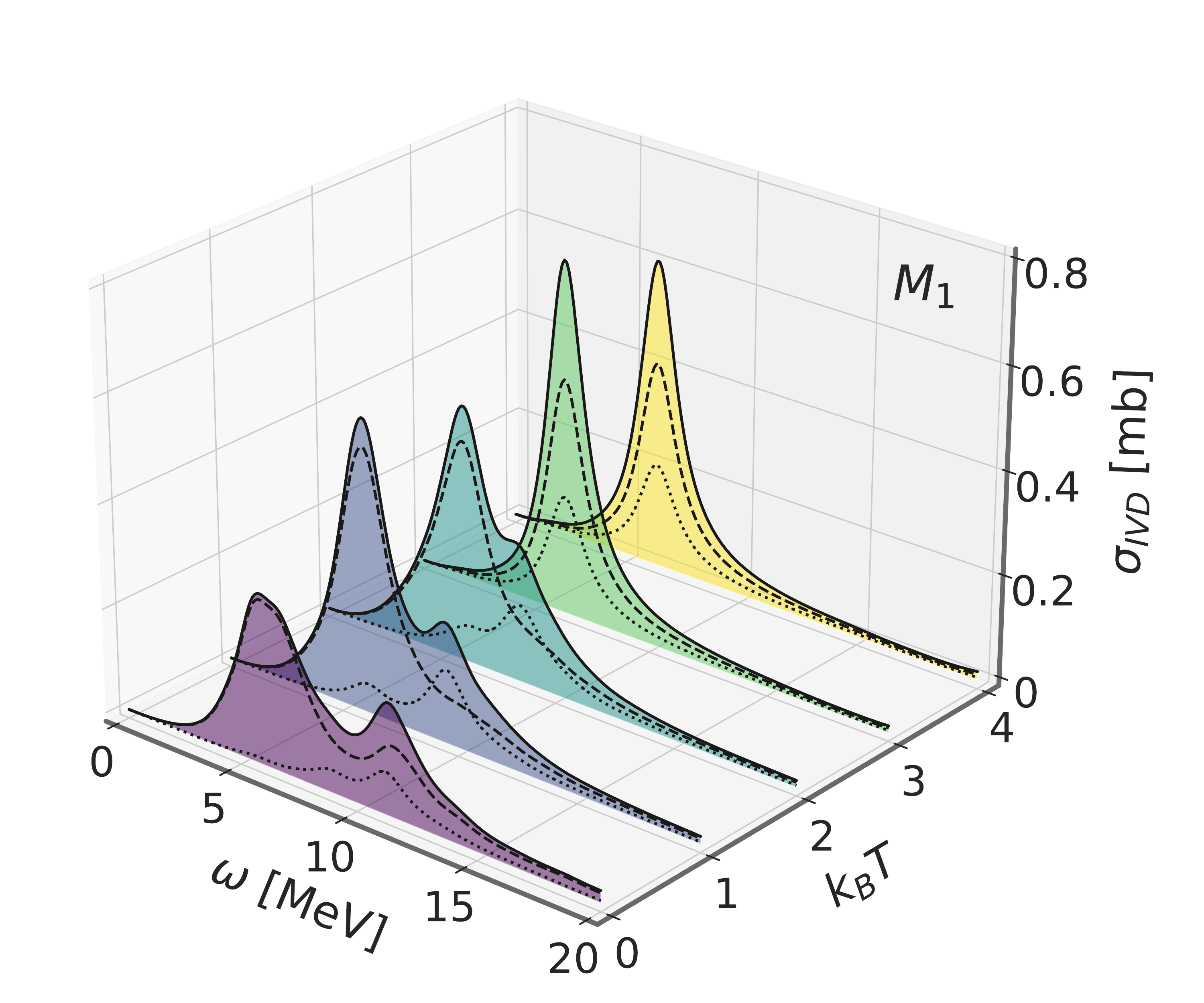}
    \caption{FT IVD electric ($E1$) (top) and magnetic ($M1$) (bottom) components of integrated photoabsorption cross section in $^{56}$Fe.}
    \label{fig:ivd_3dplot}
\end{figure}

\paragraph{FT IVD excitations. --}
Next, temperature effects are investigated in $^{56}$Fe by displaying in Fig.~\ref{fig:ivd_3dplot} the evolution of the integrated IVD photoabsorption cross section, decomposed into its angular-momentum $K=0$ and $|K|=1$ electric and magnetic components\footnote{Due to the axial and time-reversal symmetry imposed in the HFB calculation, $K=1$ and $K=-1$ HFB-QRPA responses contribute identically.}, over the temperature interval  $k_B T \in [0, 4]$\,MeV.

At ZT, the maxima of the $K=0$ and $|K|=1$ components are shifted apart as a result of the intrinsic deformation of $^{56}$Fe ground-state. While both $|K|$ components are equally important in the electric response, the $|K|=1$ component largely dominates the magnetic one.

\begin{figure}[t!]
    \centering
    \includegraphics[width=\linewidth]{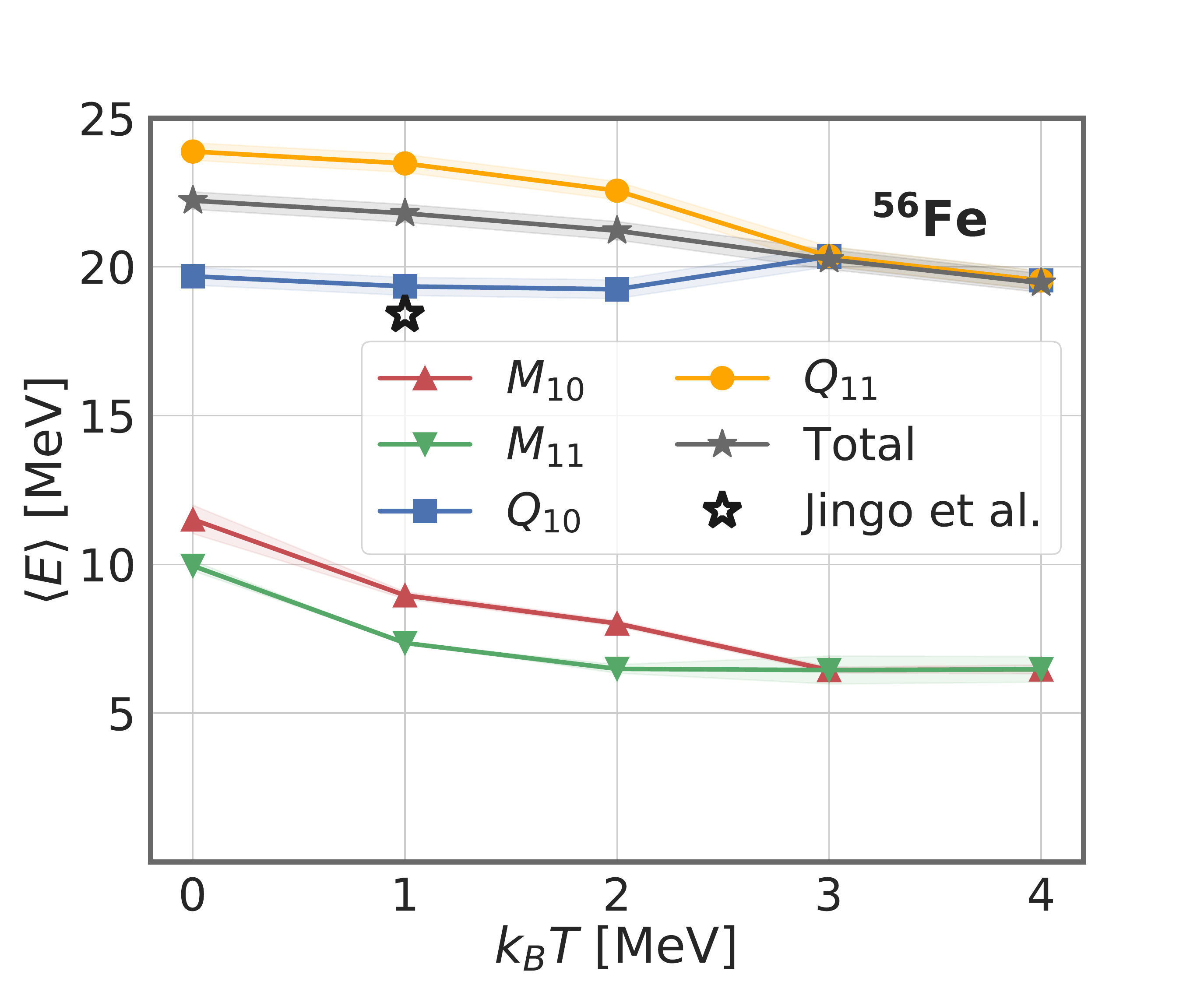}
    \caption{Thermal evolution of mean excitation energies of the different dipoles modes in $^{56}$Fe. The white star denotes the experimental measurement of Ref.~\cite{jin18}.}
    \label{fig:56Fe_Emean}
\end{figure}

Increasing the temperature induces two different processes. On the one hand, the temperature smears out the Fermi-Dirac HFB quasi-particle occupation factors compared to the ZT mean-field calculation, with the effect of increasing the matter radius $R$ of the system. In schematic models of the GDR in spherical nuclei by Goldhaber and Teller (GT)~\cite{gol48} or by Steinwedel, Jensen and Jensen (SJJ)~\cite{ste50}, this effect induces a decrease of the peak energy according to $R^{-1/2}$ (GT) and $R^{-1}$ (SJJ) laws, respectively. This effect is indeed qualitatively visible for the $|K|=1$ component of the electric response in Fig.~\ref{fig:ivd_3dplot} and more clearly illustrated in Fig.~\ref{fig:56Fe_Emean}. On the other hand, the temperature generates a collective transformation of the mean-field at play, driving the system from being intrinsically deformed at ZT to being spherical above a critical temperature $T_c$. Because the $K=0$ electric component operates along the symmetry axis, the change from prolate to spherical shape leads to an effective shrinking of the matter distribution in this direction. As a result of these two competing effects, the $E_{10}$ mean energy undergoes almost no evolution up to $T_c$ where it eventually merges with the decreasing $E_{11}$ mean energy, initially located at higher energy. The merging beyond $T_c$ of the initially different $K=0$ and $|K|=1$ responses visible in Fig.~\ref{fig:ivd_3dplot} constitutes the fingerprint of the phase transition associated with the restoration of spherical symmetry induced by the temperature. Beyond that point, the main resonance keeps evolving downwards, whereas the increase of thermal excitations enhances the dipole strength at $\omega\lesssim 12$\,MeV. 

Magnetic modes being located at much lower energies than electric ones, they bear greater sensitivity to thermal excitations that dominate their evolution. As a result, the mean $K=0$ and $|K|=1$ excitation energies continuously decrease until their merging at $T_c$. 

Eventually, the IVD response is mostly driven by the electric modes, resulting in a mean excitation energy of about 22 MeV at $k_BT=1$ MeV. This value is higher than the experimental centroid, located at 18.4 MeV. Present results indicate that  uncertainties associated with the chiral expansion of the nuclear Hamiltonian and the truncation of the computational basis are not responsible for this 20\% discrepancy. While one may inquire the error associated with discarded terms beyond three-body operators in the SRG evolution of the Hamiltonian, dynamical correlations beyond HFB-QRPA are most likely responsible and must thus be considered in ab initio FT calculations in the future.

\begin{figure}[t!]
    \centering
    \includegraphics[width=\linewidth]{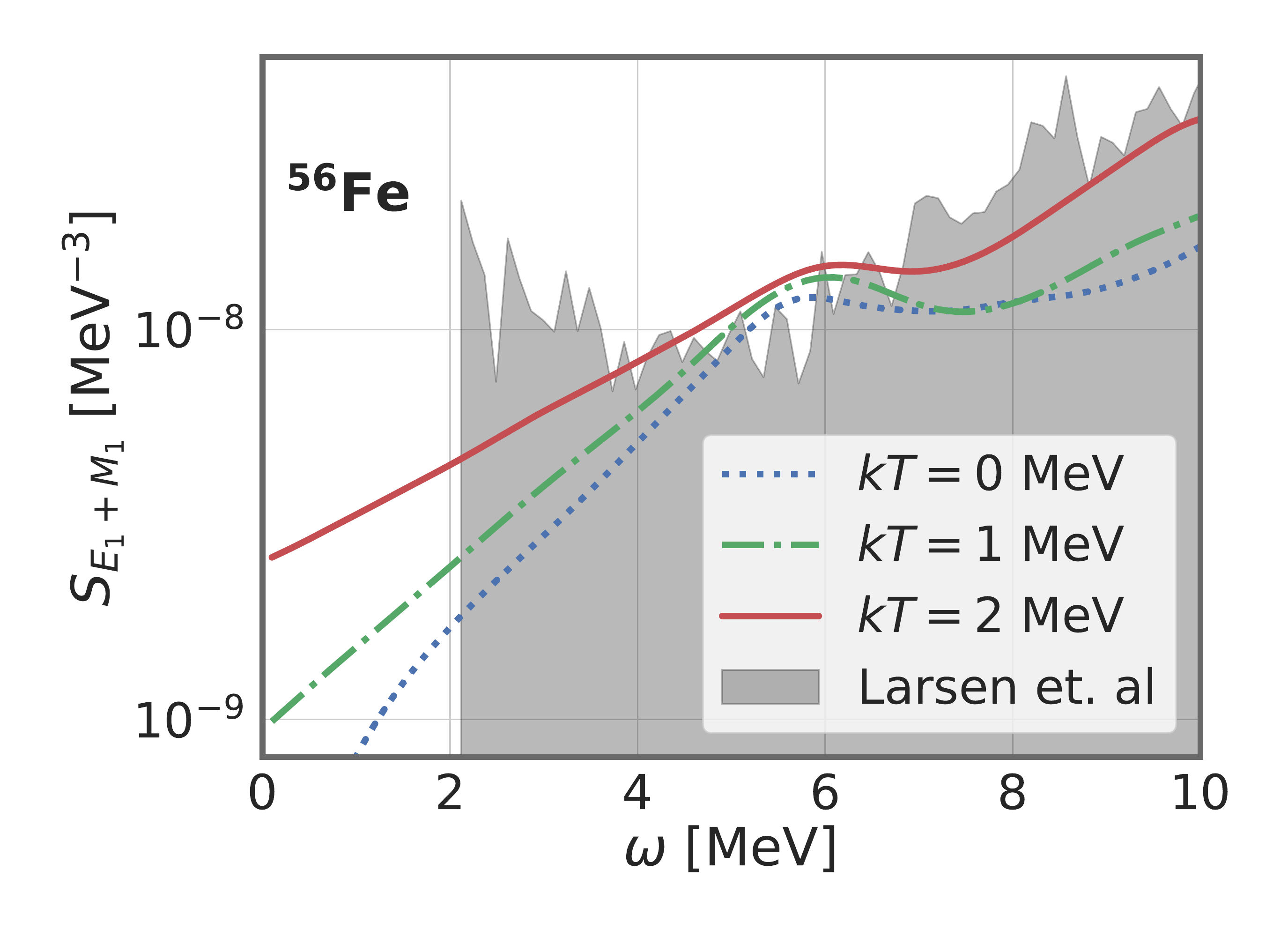}
    \caption{Low-lying FT HFB-QRPA total electro-magnetic (\(E_1+M_1\)) response in \({}^{56}\)Fe as a function of the excitation energy \(\omega\). Experimental strength-function are extracted from~\cite{Lar13}.}
    \label{fig:strength_low}
\end{figure}

The low-energy part ($\omega \leq 10$\,MeV) of measured and predicted total electromagnetic $(E1+M1)$  IVD response functions are displayed in Fig.~\ref{fig:strength_low}. The experimental $\gamma-$strength functions~\cite{Lar13} were extracted using the Oslo method~\cite{SCHILLER2000498_Oslo,midtbo2021new_Oslo}, from which the temperature of the initial state was estimated to be approximately $k_BT=1$ MeV. Experimentally, the Oslo method focuses on the de-excitation strength function, whose theoretical description still constitutes an active field of research \cite{Sieja17,Goriely18}. The FT de-excitation strength-function can be obtained from the photoabsorption strength function computed via FT HFB-QRPA by correcting the latter with the multiplicative factor \((1-e^{-\frac\omega{kT}})^{-1}\), effectively enhancing the low-lying part of the strength~\cite{Wibowo19}.

As visible from Fig.~\ref{fig:strength_low}, the experimental dipole response displays an upbend towards $\omega=0$\, MeV. The upbend, particularly pronounced in \({}^{56}\)Fe, constitutes a phenomenon of upmost interest, especially given that it is expected to significantly impact thermal neutron capture cross-sections~\cite{Brown14}. While temperature effects successfully prevent the collapse of the de-excitation theoretical strength function at low energy, the experimental trend of the upbend is not fully reproduced by the FT HFB-QRPA calculation. The origin of this discrepancy, which could be due to other multipolarities actually contributing non-negligibly to the experimental strength, to missing dynamical correlations beyond HFB-QRPA and/or to the nuclear Hamiltonian itself, will have to be investigated in future works.

\paragraph{Conclusions. --} This letter presents the first ab initio description of electromagnetic response functions at FT in mid-mass nuclei with a method handling simultaneously pairing correlations and deformation, i.e., allowing the indiscriminate study of doubly closed-shell, singly open-shell and doubly open-shell nuclei. 

After demonstrating in $^{16}$O that the ab initio HFB-QRPA constitutes a viable approach to electromagnetic responses in nuclei, the numerically-affordable QFAM implementation is employed on the basis of two- and three-nucleon interactions derived from a low-energy effective theory of QCD to investigate ZT and FT IVD photoabsorption cross sections in the doubly open-shell $^{28}$Si and $^{46}$Ti nuclei. After obtaining an excellent account of experimental data at ZT, the impact of increasing the temperature is scrutinized in \({}^{56}\)Fe. One must of course keep in mind that quantities sensitive to the details of the strength, such as the dipole polarizability provided in Tab.~\ref{alphaD} for the nuclei of present interest, require the inclusion of dynamical correlations beyond HFB-QRPA to reach fully converged values.

The numerical tool presently introduced and the first results obtained with it open the path to systematic ab initio calculations of nuclear responses to electroweak probes at ZT and FT across a large portion of the nuclear chart. 

\paragraph{Acknowledgements. --} We would like to thank S. Goriely for interesting discussions and for providing us with experimental strength functions, A. Porro for cross-checking our calculations, and N. Dubray for his contribution to the implementation of the deformed HFB solver. Calculations were performed using HPC resources from GENCI-TGCC (Contracts No. A0090507392 and A0110513012).
RR acknowledges support by the DFG through SFB 1245 (Project ID 279384907) and the BMBF through Verbundprojekt 05P2021 (ErUM-FSP T07, Contract No. 05P21RDFNB). 
\bibliography{main.bib}

\end{document}